\documentclass[]{rcdj}

\begin{document}

\journal{REGULAR AND CHAOTIC DYNAMICS, V.\,8, \No2, 2003}

\setcounter{page}{201}

\received 22.08.2002.

\amsmsc{37J60, 37J35, 70G45}

\doi{10.1070/RD2003v008n02ABEH000237}

\title{DYNAMICS OF ROLLING DISK}

\runningtitle{DYNAMICS OF ROLLING DISK} \runningauthor{A.\,V.\,BORISOV,
I.\,S.\,MAMAEV, A.\,A.\,KILIN}

\authors{A.\,V.\,BORISOV, I.\,S.\,MAMAEV, A.\,A.\,KILIN}
{Institute of Computer Science\\
Universitetskaya, 1\\
426034, Izhevsk, Russia\\
E-mail: borisov@rcd.ru\\
E-mail: mamaev@rcd.ru\\
E-mail: aka@rcd.ru}

\abstract{In the paper we present the qualitative analysis of rolling
motion without slipping of a homogeneous round disk on a horisontal plane.
The problem was studied by S.\,A.\,Chaplygin, P.\,Appel and D.\,Korteweg who
showed its integrability. The behavior of the point of contact on a plane
is investigated and conditions under which its trajectory is finit are
obtained. The bifurcation diagrams are constructed.
}

\maketitle
\section{Introduction}
{\advance\baselineskip by -.5pt
For the first time the motion of a heavy dynamically symmetrical round
disk on a horizontal absolutely rough plane was investigated by G\,Slesser
(1861)~\cite{Slesser}, N.\,Ferrers (1872)~\cite{Ferrers}, K.\,Neumann
(1886), and A.\,Firkandt (1892). These studies eventually (after
unsuccessful attempts by Neumann and Lindel{\"o}f) lead to the correct
form of equations of motion. This form differs from the usual (Lagrangian
or Hamiltonian) equations of mechanics because of the nonholonomic
constrain showing that the velocity of the point of contact of a disk with
a plane is zero. We shall not discuss in detail the general forms of the
equations of the nonholonomic mechanics (they are presented, for example,
in~\cite{Markeev},~\cite{NeimFuf}). Instead, we concentrate on the pretty obvious
form of these equations obtained from the general principle of
dynamics~--- the conservation law of the moment of momentum written in the
disk-fixed axes.

S.\,A.\,Chaplygin (1897) was the first to show the integrability of the
problem on rolling motion of a disk. He presented the reduction of the
problem to the analysis of hypergeometric quadratures in
paper~\cite{Chap}, where he showed also the integrability of the problem
on rolling motion of an arbitrary heavy dynamically symmetric body of
rotation on a horizontal plane~--- in the latter case the problem is
reduced to the integration of the linear differential second-order
equation. The integration of equations of motion of a disk in
hyperelliptic functions was also performed in 1900 independently from each
other and from Chaplygin by P.\,Appel~\cite{Appel}
and~D.\,Korteweg~\cite{Korteweg}. Sometimes the problem on rolling motion
of a disk is referred to as Appel--Korteweg problem (or simply Appel
problem), but this is, probably, not quite correct. In 1903 the same
result has been rediscovered by E.\,Gellop~\cite{Gellon}, however he used
the Legendre functions.

Despite of the explicit hypergeometric quadratures the various qualitative
properties of disk motion were not studied for the long time. There were
mainly studies of stationary motions and of their stability (the
corresponding bibliography is presented in book~\cite{Markeev}). Some
qualitative properties of the disk motion have been discussed only in
papers S.\,N.\,Kolesnikov~\cite{Kolesnikov} and
Yu.\,N.\,Fedorov~\cite{Fedorov}. The first paper shows that for almost all
initial conditions the disk never falls onto a plane and the second one
present the procedure of investigation of the reduced system. Analogous
results for the dynamically asymmetrical disk and disk moving on an
inclined plane (nonintegrable problems) were obtained in~\cite{Afonin},~\cite{KozDvizh}.
Among the modern works analyzing the rolling motion of the disk
we shall note papers O.\,M.\,O'Reily~\cite{OReily}, R.\,Cushman,
J.\,Hermans, D.\,Kemppainen~\cite{KushHerKem}, and A.\,S.\,Kuleshov
\cite{Kuleshov} devoted to the study of bifurcations and stability of
stationary motions of the disk. \looseness=1

General results of a qualitative analysis for the rolling motion of a
heavy body of rotation were obtained in paper
N.\,K.\,Moshuk~\cite{Moshuk}. The paper include the frequency analysis,
application of the KAM-theory, and basic qualitative properties for the
motion of the point of contact. It appears that the point of contact
performs the composite bounded motion: it periodically traces some closed
curve which rotates as a rigid body with some constant angular velocity
about the fixed point. Thus the realization of some resonance relation
between frequencies makes possible the drift of the body of rotation to
the infinity.

In this paper we develop these qualitative considerations and complement
them with the computer analysis. We also present various types of
trajectories which are traced by the point of contact in the body-fixed
and relative frames of references since they have curious forms which are
difficult to predict. Using the computer modelling we explicitly
investigate the hypothesis about the drift to the infinity under the
resonance conditions. We present the most general three-dimensional
bifurcation diagram in the space of the first integrals and the complete
atlas of its sections by various planes, constructed with the help of
computer modelling. \looseness=1

In this paper we also present a new method of reduction of the problem to
an one-degree integrable Hamiltonian system and explicitly consider the
existence of Hamiltonian formulation for different variants of equations
of motion of the problem.}

\section{The rolling motion of a rigid body on a plane}

\subsection{Equations of motion and their integrals}

Let the rigid body in an exterior field of force perform a rolling motion
on a plane without sliding. In this case the equations of motion have the
most convenient form in the body-fixed frame of references which axes are
directed along the principal axes of inertia of the body and the origin is
situated at the center of mass. In the following text all vectors are
assumed to be projected on these axes.

\wfig<bb=0 0 53.5mm 38.3mm>{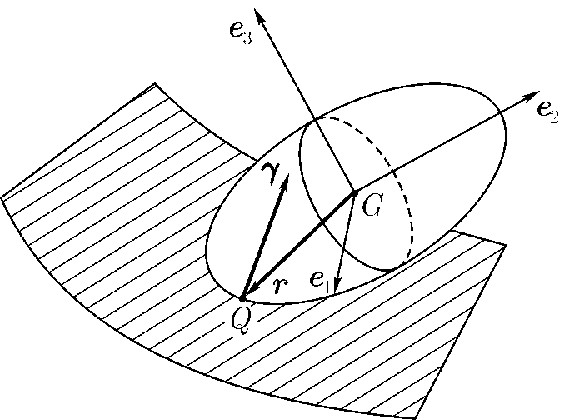}

The condition of absence of slipping thus becomes \eqc[eq0]{
\bv+\bs\omega\x\br=0, } where $\bv,\,\bs\omega$ are the velocity of the
center of mass and the angular velocity of the body and $\br$ is the
vector directed from the center of mass to the point of contact
(see~fig.~\ref{pic1.eps}).

Let's denote the projections of the fixed basis vectors to the moving axes
by $\bs\alpha,\,\bs\beta,\,\bs\gamma$ (the vector $\bs\gamma $ is
perpendicular to the plane) and by $(x,\,y)$ we shall denote the
coordinates of the projection of the center of mass onto the plane in the
fixed frame of references. We assume that the field of force is potential
with a potential depending only on the orientation of the body
$U=U(\bs\alpha,\,\bs\beta,\,\bs\gamma)$. The complete set of the equations
of motion defining the given system can be represented in the form
\vspace{-1mm}\eqnc{
\label{eq1.1}\dot{\bM}=\bM\x\bs\omega+m\dot{\br}\x(\bs\omega\x\br)+\bs\alpha\x\pt{U}{\bs\alpha}+
\bs\beta\x\pt{U}{\bs\beta}+\bs\gamma\x\pt{U}{\bs\gamma},\\
\label{eq1.2}\dot{\bs\alpha}=\bs\alpha\x\bs\omega,\qq
\dot{\bs\beta}=\bs\beta\x\bs\omega,\qq
\dot{\bs\gamma}=\bs\gamma\x\bs\omega. } The equation (\ref{eq1.1})
describes the evolution of the vector of moment of momentum for the body
with respect to the point of contact $\bM$ and (\ref{eq1.2}) concerns the
evolution of the fixed basis vectors in the body-fixed frame of
references.

The motion of the center of mass can be obtained in quadratures from
solutions of the equations~(\ref{eq1.1}), (\ref{eq1.2}) as follows
\vspace{-1mm}
\eqc[k-eq1.3]{ \dot{x}=(\br\x\bs\omega,\,\bs\alpha),\qq
\dot{y}=(\br\x\bs\omega,\,\bs\beta). \vspace{-1mm}}

The expression of the vector of moment of momentum with respect to the
point of contact $\bM$ can be written in the following form \eqc[eq2]{
\bM={\bf I}\bs\omega+m\br\x(\bs\omega\x\br), } where ${\bf
I}=\diag(I_1,\,I_2,\,I_3)$ is the tensor of inertia of the body. In turn
$\br$ can be uniquely expressed (for a convex body) through the normal to
the plane $\bs\gamma$ from the equation
\vspace{-1mm}
\eqc[eq3]{ \bs\gamma=-\frac{\nb
F(\br)}{|\nb F(\br)|}, \vspace{-1mm}} Here $F(\br)=0$ is the equation of the body's
surface.

Let's consider a motion of the point of contact on a plane. If we denote
the position of the point of contact on the plane in the fixed frame of
references as~$(X,\,Y)$, then the equation of motion for the point of
contact can be presented in the form
\vspace{-1mm}
\eqc[eq7]{
\dot{X}=(\dot{\br},\bs\alpha),\qq\dot{Y}=(\dot{\br},\bs\beta)\,. \vspace{-1mm}} where
$\dot{\br}$ is determined from equations (\ref{eq1.1})~-- (\ref{eq3}).
Actually $\dot{X}$ and $\dot{Y}$ are projections of the velocity of the
point of contact in the relative frame of reference onto the fixed axes.

The equations of motion in the form similar to~(\ref{eq1.1})~--
(\ref{eq1.2}) are presented, for example, in book \cite{Karapetyan}. They
can be obtained also by means of Poincar{\'e}\f Chetaev formalism \cite
{BorMamDTT} with undetermined Lagrangian coefficients; these coefficients
shall be eliminated with the help of the constrains'
equations~(\ref{eq0}).

The system~(\ref{eq1.1})~-- (\ref{eq1.2}) generally has seven independent
integrals of motion, six of them are trivial geometrical integrals:
\vspace{-1mm}
\eqc[eq5]{
{\bs\alpha}^2=1,\qq{\bs\beta}^2=1,\qq{\bs\gamma}^2=1,\\
(\bs\alpha,\,\bs\beta)=0,\qq(\bs\beta,\,\bs\gamma)=0,\qq(\bs\gamma,\,\bs\alpha)=0.\vspace{-1mm}
} The seventh is the integral of energy
\vspace{-1mm}
\eqc[eq6]{
\frac{1}{2}(\bM,\,\bs\omega)+U(\bs\alpha,\,\bs\beta,\,\bs\gamma)=h=\const.\vspace{-1mm}
} Generally the given system has no other additional integrals and the
possibility of its integrability in concrete cases depends on the presence
of additional tensor invariants (measure, fields of symmetry, integrals).

\subsection{The rolling motion of a heavy disk}

Let's consider the case of rolling motion for an axially symmetric disk of
radius $R$ in the field of gravity. The field is, obviously, also axially
symmetric with the potential depending only on $\bs\gamma$. Moreover, we
suppose that the disk is dynamically symmetric, i.~e. $I_1=I_2 $. The
potential energy in this case has the following form
\vspace{-1mm}
\eq[712]{
U=-mg(\br,\,\bs\gamma)=mgR\sqrt{1-\gamma_3^2}.
\vspace{-1mm}}

The equation of surface for the disk is $F(\br)=r_1^2+r_2^2-R^2$.
Substituting it in the equation~(\ref{eq3}) and solving with respect to
$\br$ we obtain
\vspace{-1mm}
\eq[eq4]{ r_1=-\frac{R\gamma_1}{\sqrt{1-{\gamma}^2_3}},\qq
r_2=-\frac{R\gamma_2}{\sqrt{1-{\gamma}^2_3}},\qq r_3=0.
\vspace{-1mm}}

As the potential energy depends only on $\bs\gamma$, in the equations of
motion (\ref{eq1.1})~-- (\ref{eq1.2}) we get the separate system of six
equations. \eqc[eq8]{
\dot{\bM}=\bM\x\bs\omega+m\dot{\br}\x(\bs\omega\x\br)+mg\br\x\bs\gamma,\\
\dot{\bs\gamma}=\bs\gamma\x\bs\omega. } Expressing $\bs\omega,\,\br$ from
relations~(\ref{eq2}), (\ref{eq4}) we shall get the closed system for the
variables $\bM,\,\bs\gamma$ similar in many aspects is to the
Euler--Poisson system in the Lagrange case, however the obtained system is
much more complicated than the last one.

The equations~(\ref{eq8}) preserve the geometrical integral $\bs\gamma^2$
and the energy~(\ref{eq6}), in addition they allows the standard invariant
measure (with a constant density). For the integrability (by
Euler-Jacobi~\cite{KozlovMetody}) of these equations we need two
additional integrals. In the following we describe the method of
derivation of these integrals.

The possibility of separation of the system~(\ref{eq8}) from the general
system (\ref{eq1.1})~-- (\ref{eq1.2}) is connected to the symmetry with
respect to the rotations about the vertical axis defined by the vector
$\bs\gamma$. The system~(\ref{eq8}) is invariant with respect to the field
of symmetries commuting with the vector field of the problem. \eqc[eq812]{
\widehat{\bv}_{\psi}=\alpha_1\pt{}{\beta_1}-\beta_1\pt{}{\alpha_1}+
\alpha_2\pt{}{\beta_2}-\beta_2\pt{}{\alpha_2}+
\alpha_3\pt{}{\beta_3}-\beta_3\pt{}{\alpha_3}, } It is possible to show
that the variables $\bM,\,\bs\gamma$ are the integrals of
field~(\ref{eq812}) that is $\widehat{\bv}_{\psi}(M_i)=0$,
$\widehat{\bv}_{\psi}(\gamma_i)=0$, $i=1,\,2,\,3$. According to the
general Lie theory~\cite{teorLi}, variables $\bM,\,\bs\gamma$ define the
reduced system. For the classical Euler--Poisson equations the
corresponding reduction is the Raus reduction with respect to the cyclical
angle of precession.

In addition to the field of symmetries~(\ref{eq812}) the equations of
motion~(\ref{eq1.1})~-- (\ref{eq1.2}) for the axially symmetric body allow
one more field of symmetries corresponding to the rotation about the axis
of symmetry of the disk. \eqc[eq9]{
\widehat{\bv}_{\vfi}=M_1\pt{}{M_2}-M_2\pt{}{M_1}+\gamma_1\pt{}{\gamma_2}-\gamma_2\pt{}{\gamma_1}+\\
+\alpha_1\pt{}{\alpha_2}-\alpha_2\pt{}{\alpha_1}+\beta_1\pt{}{\beta_2}-\beta_2\pt{}{\beta_1}.
}

It is possible to show that integrals of the field~(\ref{eq9}) are
projections of the moment and normal to the plane of disk onto the fixed
axes of coordinates \eqc*{
\bN=((\bM,\,\bs\alpha),\,(\bM,\,\bs\beta),\,(\bM,\,\bs\gamma)),\qq
\bn=(\alpha_3,\,\beta_3,\,\gamma_3). } The equations of motion for these
variables can be presented in the following form \eqc[eq912]{
\dot{\bN}=m\dot{\widetilde{\br}}\x(\widetilde{\bs\omega}\x\widetilde{\br})+mg\widetilde{\br}\x\bn,\\
\dot{\bn}=\widetilde{\bs\omega}\x\bn, } where symbols
$\widetilde{\bs\omega},\,\widetilde{\br}$ denote the same vectors, but
projected onto the fixed axes (that is $\widetilde{\omega}_1 =
(\bs\omega,\,\bs\alpha),\ldots,\,\widetilde{r}_1 =
(\br,\,\bs\alpha),\ldots $). The explicit expression of the components of
the vector $\widetilde{\br}$ is \eqc[934]{
\widetilde{\br}=\left(\frac{R\alpha_3\gamma_3}{\sqrt{1-\gamma_3^2}},\,\frac{R\beta_3\gamma_3}{\sqrt{1-\gamma_3^2}},\,-R\sqrt{1-\gamma_3^2}\right).
} The vector $\bN$ is expressed through $\bs\omega$ by the formula
\eqc[eq978]{
N=I_1\widetilde{\bs\omega}+(I_3-I_1)(\widetilde{\bs\omega},\,\bn)\bn+
m\widetilde{\br}\x(\widetilde{\br}\x\widetilde{\br}). }

\begin{rem}
\label{rem2} Such reduction is also possible for an arbitrary body of
rotation.
\end{rem}

\subsection{A reduction to the integrable one-degree Hamiltonian system}

Let's describe the process of reduction of order with respect to the both
fields of symmetries~(\ref{eq812}) and (\ref{eq9}). For that we shall
choose the simultaneous integrals of these fields as variables of the
reduced system. According to~\cite{BorMamInvMes}, the most convenient
algebraic set of such variables is \nopagebreak\vspace{-3mm} \eqc[eq10]{
\gamma_3,\q K_1=M_1\gamma_1+M_2\gamma_2=N_3-\gamma_3(\bN,\,\bn),\\
K_2=\sqrt{\frac{I_1}{I_3+mR^2}}M_3=\sqrt{\frac{I_1}{I_3+mR^2}}(\bN,\,\bn),\\
K_3=\gamma_1M_2-\gamma_2M_1=N_1n_2-N_2n_1. } The equations of motion in
the new variables become \vspace{-2mm} \eqc[eq11]{
\dot{\gamma}_3=\frac{K_3}{I_1+mR^2},\\
\dot{K}_1=-\frac{I_3}{(I_1+mR^2)\sqrt{I_1(I_3+mR^2)}}K_3K_2,\\
\dot{K}_2=-\frac{mR^2}{(I_1+mR^2)\sqrt{I_1(I_3+mR^2)}}\frac{K_3K_1}{1-\gamma_3^2},\\
\dot{K}_3=-\frac{\gamma_3}{1-\gamma_3^2}\left(\frac{K_1^2}{I_1}+
\frac{K_2^2}{I_1+mR^2}\right)+\\
+\frac{\sqrt{I_1(I_3+mR^2)}}{I_1^2}K_1K_2+mgR\gamma_3\sqrt{1-\gamma_3^2}.
} The equations~(\ref{eq11}) preserve the invariant measure with density
$\rho=\frac{1}{1-\gamma_3^2}$. Dividing the second and the third equations
on the first and choosing a new independent variable, the angle of
nutation $\theta =\arccos\gamma_3$, we shall get the system of linear
equations \vspace{-2mm} \eqc[eq12]{
\frac{dK_1}{d\theta}=\frac{I_3\sin\theta}{\sqrt{I_1(I_3+mR^2)}}K_2,\q
\frac{dK_2}{d\theta}=\frac{mR^2}{\sqrt{I_1(I_3+mR^2)}}\frac{K_1}{\sin\theta}.
}

The general solution of these equations can be presented in the
form~\cite{Markeev}
\vspace{-2mm}\eqc[eq13]{
K_1=C_1\frac{I_3\sin^2\theta}{2\sqrt{I_1(I_3+mR^2)}}F(1+\xi,\,1+\eta,\,2,\,\frac{1-\cos\theta}{2})-\\
-C_2\frac{I_3\sin^2\theta}{2\sqrt{I_1(I_3+mR^2)}}F(1+\xi,\,1+\eta,\,2,\,\frac{1+\cos\theta}{2}),\\
K_2=C_1F(\xi,\,\eta,\,1,\,\frac{1-\cos\theta}{2})+C_2F(\xi,\,\eta,\,1,\,\frac{1+\cos\theta}{2}),
\vspace{-1mm}} where $\xi$ and $\eta$ are the solutions of the quadratic equation
$x^2-x +\frac{I_3mR^2}{I_1 (I_3+mR^2)}=0 $ and $F(\xi,\,\eta,\,n,\,z)$ is
the generalized hypergeometric function representable by series
\vspace{-2mm}
\eq[eq14]{
F(\xi,\,\eta,\,n,\,z)=\sum\limits_{k=0}^{\infty}\frac{\Gamma(\xi+k)\Gamma(\eta+k)\Gamma(n)}
{\Gamma(\xi)\Gamma(\eta)\Gamma(n+k)}\frac{z^k}{k!} \vspace{-1mm}} Thus, the
relations~(\ref{eq13}) define (implicitly) the integrals of motion. In
this case they are the ``constants'' $C_1 $ and $C_2 $ expressed through
$K_1,\,K_2,\,\theta$.

The quadrature for the angle of nutation can be obtained from the integral
of energy written in the variables $K_1,\,K_2,\,K_3,\,\theta $
\vspace{-2mm} \eqc[eq15]{
\dot{\theta}^2=2\sin^2\theta(I_1+mR^2)P(\theta),\\
P(\theta)=h-\frac{K_1^2}{2I_1\sin^2\theta}-\frac12\frac{K_2^2}{I_1}-mgR\sin\theta.
\vspace{-1mm}} Here we assume that the variables $K_1, \, K_2 $ are expressed through
the constants of integrals and angle~$\theta$ according to the
formulas~(\ref{eq13}). In this case the function $P(\theta)$ (depending on
the constants of integrals) define the analog of gyroscopic function for
the Lagrange top~\cite{BorMamDTT},~\cite{Magnus}.

\fig<bb=0 0 132.3mm 42.2mm>{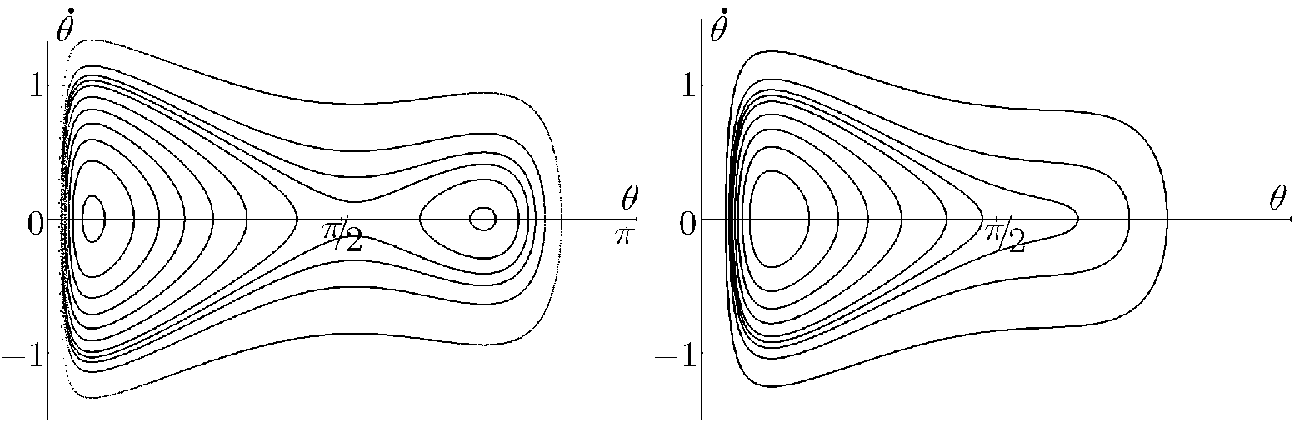}[\label{pic2} Phase portraits of the
system~(\ref{eq15}) at various values $C_1$ and $C_2$. Left: the case of
existence of three periodic solutions ($C_1=0.05,\,C_2=0.01$). Right: the
case of existence of one periodic solution ($C_1=0.08,$ ${C_2=-0.02}$).]

Thus, the equation~(\ref{eq15}) at the fixed values $C_1$ and $C_2$ define
the one-degree Hamiltonian system. The phase portraits of this system on
the plane $\theta,\,\dot{\theta}$ are presented in fig.~\ref{pic2}. All
the variables $\gamma_3,\,K_1,\,K_2,\,K_3$ are periodic functions of time
with the period $T_{\theta}$ and the corresponding frequency
$\omega_{\theta}$.

\begin{rem}
\label{rem3} According to~\cite{BorMamInvMes}, the system~(\ref {eq11}) is
the Hamiltonian one with degenerated Poisson bracket which has two Casimir
functions expressed through hypergeometric functions.
\end{rem}

\subsection{Quadratures for angles of proper rotation and a precession}

According to the general Lie theory~\cite{teorLi}, if the variables of
reduced system~(\ref{eq10}) are the given functions of time, then all the
variables of initial system~(\ref{eq8}) (and accordingly~(\ref{eq912}))
can be obtained by one quadrature (if fields
$\widehat{\bv}_{\psi}$~(\ref{eq812}) and $\widehat{\bv}_{\phi}$
(\ref{eq9}) are commuting).

Indeed, using the equalities $\tg\vfi=\frac{\gamma_1}{\gamma_2}$ (and
correspondingly $\tg\psi=-\frac{n_1}{n_2}$) for angles $\vfi$ and $\psi$,
we obtain \eqc[eq16]{
\dot{\vfi}=-\frac{\gamma_3}{1-\gamma_3^2}\frac{K_1}{I_1}+\frac{K_2}{\sqrt{I_1(I_3+mR^2)}},\qq
\dot{\psi}=-\frac{K_1}{I_1(1-\gamma_3^2)}. }

Thus, for each of the angles the dependence on time is defined as an
integral of a periodic function with the frequency $\omega_{\theta}$,
hence it can be presented in the form (see, for
example,~\cite{KozlovMetody},~\cite{Moshuk}) \eqc[eq17]{
\vfi=\omega_{\vfi}t+\vfi_{*}(t),\qq \psi=\omega_{\psi}t+\psi_{*}(t), }
where $\vfi_{*}(t),\,\psi_{*}(t)$ are periodic function with frequency
$\omega_{\theta}$. Moreover,~(\ref{eq16}) and~(\ref{eq17}) imply also {\em
that all the frequencies $\omega_{\theta},\,\omega_{\vfi},\,\omega_{\psi}$
depend only on the constants of the first integrals}.

\begin{figure}[ht!]
\centering
\cfig<width=82.8mm>{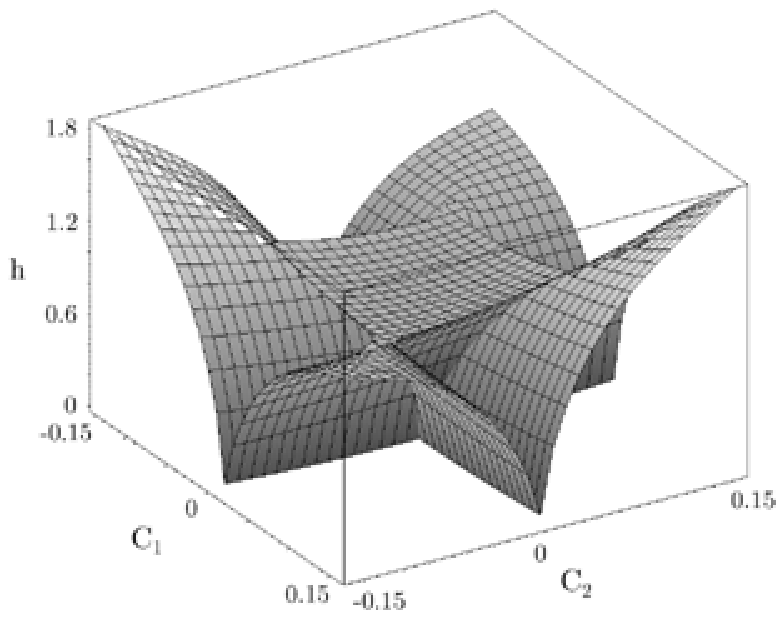}
\caption{The surface of regular precessions.
Parameters of the system are $I_1=0.25,\,I_2=0.5,\,R=1,\,m=1,$ ${g=1}$.}
\label{pic3}
\end{figure}

\subsection{A motion of the point of contact}

Following papers~\cite{KozKol},~\cite{Moshuk} we present the equation for the
velocity of the point of contact in the form
\vspace{-2mm}\eqc[eq18]{
\dot{Z}=R\left(\frac{\gamma_3}{1-\gamma_3^2}\frac{K_1}{I_1}-\frac{K_2}{\sqrt{I_1(I_3+mR^2)}}\right)e^{i\psi},
\vspace{-2mm}} where $Z=X+iY$ and $X,\,Y$ are the coordinates of the point of contact
in the fixed frame of references.

Thus the coordinates of the point of contact are determined by quadratures
of quasiperiodic two-frequency (with the frequencies
$\omega_{\psi},\,\omega_{\theta}$) functions of time.

\section{The qualitative analysis and results}

Let's perform the qualitative analysis of the dynamics of the disk motion.
We will make a classification of all possible motions depending on the
constants of the first integrals. Some features of the considered case
essentially complicate this work in comparison with the case of the
Lagrange top for Euler--Poisson equations. For uniformity we recommend to
study such analysis for the Lagrange case in book~\cite{BorMamDTT}. The
complexity of analysis is caused by the facts that the integrals of motion
can not be expressed in elementary functions (only in special one) and the
system has no natural Hamiltonian presentation. Moreover, in addition to
the motion of apexes of the body (disk) we shall classify trajectories of
the point of contact obtained by additional quadratures of quasiperiodic
functions.

\subsection{The bifurcation analysis of the reduced system}

\begin{figure}[t!]
\centering
\cfig<bb=0 0 100.0mm 56.4mm>{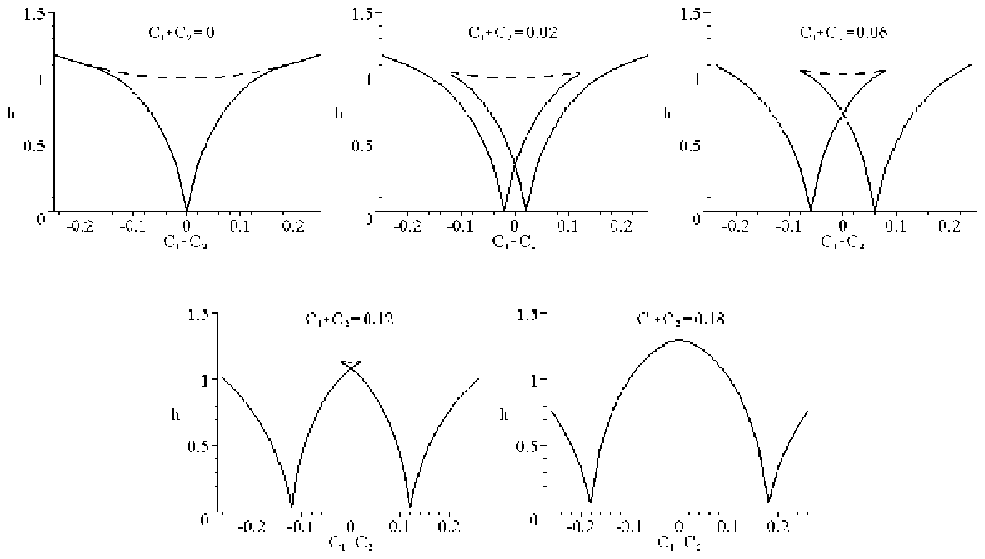}
\bigskip
\caption{Sections of the
surface of regular precessions represented in fig.~\ref{pic3} by
planes $C_1+C_2=\const$.}
\label{pic4a}
\end{figure}

\begin{figure}[t!]
\centering
\cfig<bb=0 0 100.0mm 56.3mm>{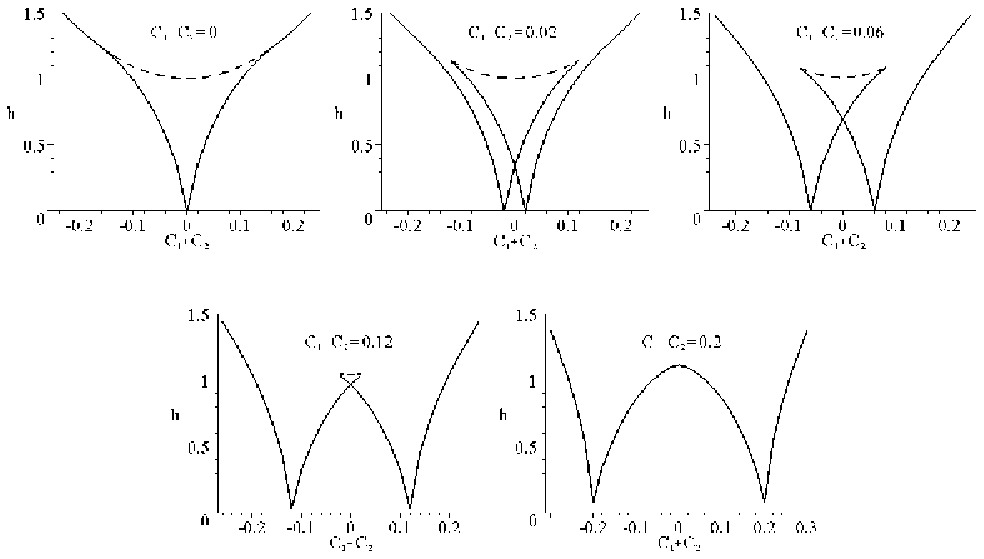}
\bigskip
\caption{Sections of the surface of regular
precessions represented in fig.~\ref{pic3} by planes
$C_1-C_2=\const$.}
\label{pic4b}
\end{figure}

{\advance\baselineskip by -0.3pt
Possible types of motion for the axis of symmetry of the body are
completely determined by the form of the gyroscopic function $P(\theta)$
and by the energy level. Critical values of the integrals of motion
$C_1,\,C_2,\,h $ are determined by the equations \eq[eq19]{
P(\theta)=0,\qq \frac{d P(\theta)}{d\theta}=0. }

\noindent In three-dimensional space with coordinates $C_1,\,C_2,\,h$
equations~(\ref{eq19}) define a three-dimensional surface, so-called {\it
surface of regular precessions}~\cite{BorMamDTT} (see
fig.~\ref{pic3}). This name is connected to the fact that at the given
values of integrals the coin performs motion with the fixed angle
$\theta=\const$, which is analogous to the precession for Lagrange
top~\cite{Magnus}. The full atlas of sections of the surface of regular
precessions (bifurcation diagrams) by planes $C_1+C_2 =\const $ and
$C_1-C_2 =\const $ is presented in figs.~\ref{pic4a} and \ref{pic4b}
accordingly. In fig.~\ref{pic5a} and~\ref{pic5b} for two different
sections we show the forms of the gyroscopic function $P(\theta)$
corresponding to various values of integrals $C_1,\,C_2,\,h$.

\fig<bb=0 0 110.0mm 64.6mm>{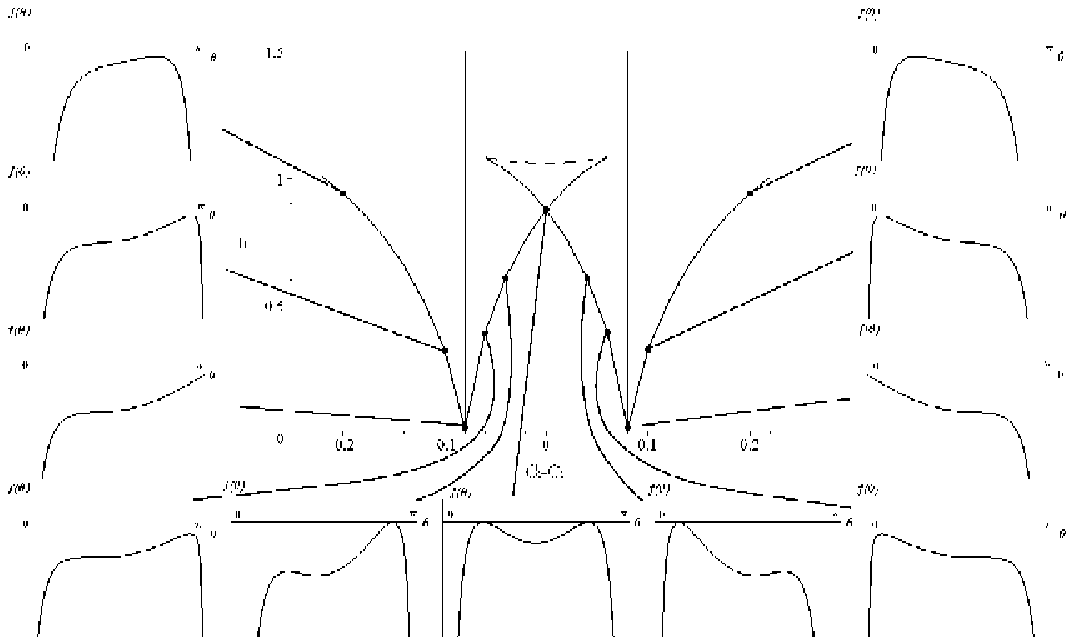}[\label{pic5a} Various types of the
gyroscopic function for the section of the surface of regular precessions
by plane $C_1+C_2=0.08$.] \fig<bb=0 0 110.0mm
64.6mm>{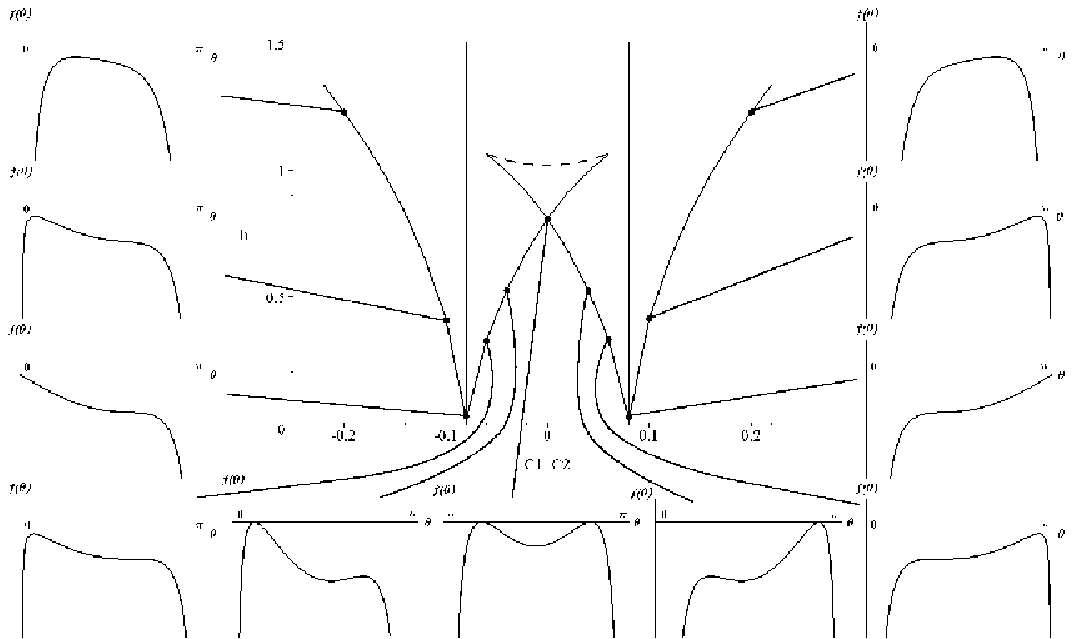}[\label{pic5b} Various types of the gyroscopic function
for the section of the surface of regular precessions by plane
$C_1-C_2=0.08$.\vspace{-3mm}]

Using these figures (and the rule of signs) we can easily study the
stability of the corresponding solutions located on branches of the
bifurcation diagram (branches corresponding to unstable solutions are
represented on the diagram by a dotted line). In figs.~\ref{pic5a}
and~\ref{pic5b} vertical straight lines represent cases when $C_1=0$ or
$C_2=0$. In these cases the disk motion corresponds to the falling and
planes determined by these equalities define in space of integrals
$C_1,\,C_2,\,h$ the two-dimensional manifold of fallings. Thus for almost
all initial conditions the disk do not fall performing the rolling motion
on a plane.

Other remarkable motions correspond to the cases $C_1=C_2 $, the rolling
motion of the disk, and $C_1 =-C_2 $, the rotation of the disk about its
axis passing through the diameter. During the latter motion the
declination of the disk with respect to the vertical remains constant.

\begin{rem}
\label{rem4} 
The bifurcation diagram (fig.~\ref{pic3}, \ref{pic4a}, \ref{pic4b}) is
different from one presented in papers~\cite{Kuleshov},~\cite{OReily} since
instead of the value of energy we use the value of angle of declination
corresponding to the precession $\theta_0$ and this function has no
physical sense for other motions (when this angle is not preserved). Only
the points on the surface of regular precessions have the physical sense.
At the same time each value of constants $C_1,\,C_2,\,h$ in space of
integrals in fig.~\ref{pic3} corresponds to some motion whether this
point is situated on the surface of regular precessions or not and this is
important for the qualitative analysis.
\end{rem}
\begin{rem}\label{rem5}
One of sections of the three-dimensional diagram by a plane $h=\const$ and
the corresponding gyroscopic functions are presented in paper
\cite{KushHerKem}.
\end{rem}
}

\begin{figure}[t!]
\begin{center}
\cfig<bb=0 0 110.0mm 136.4mm>{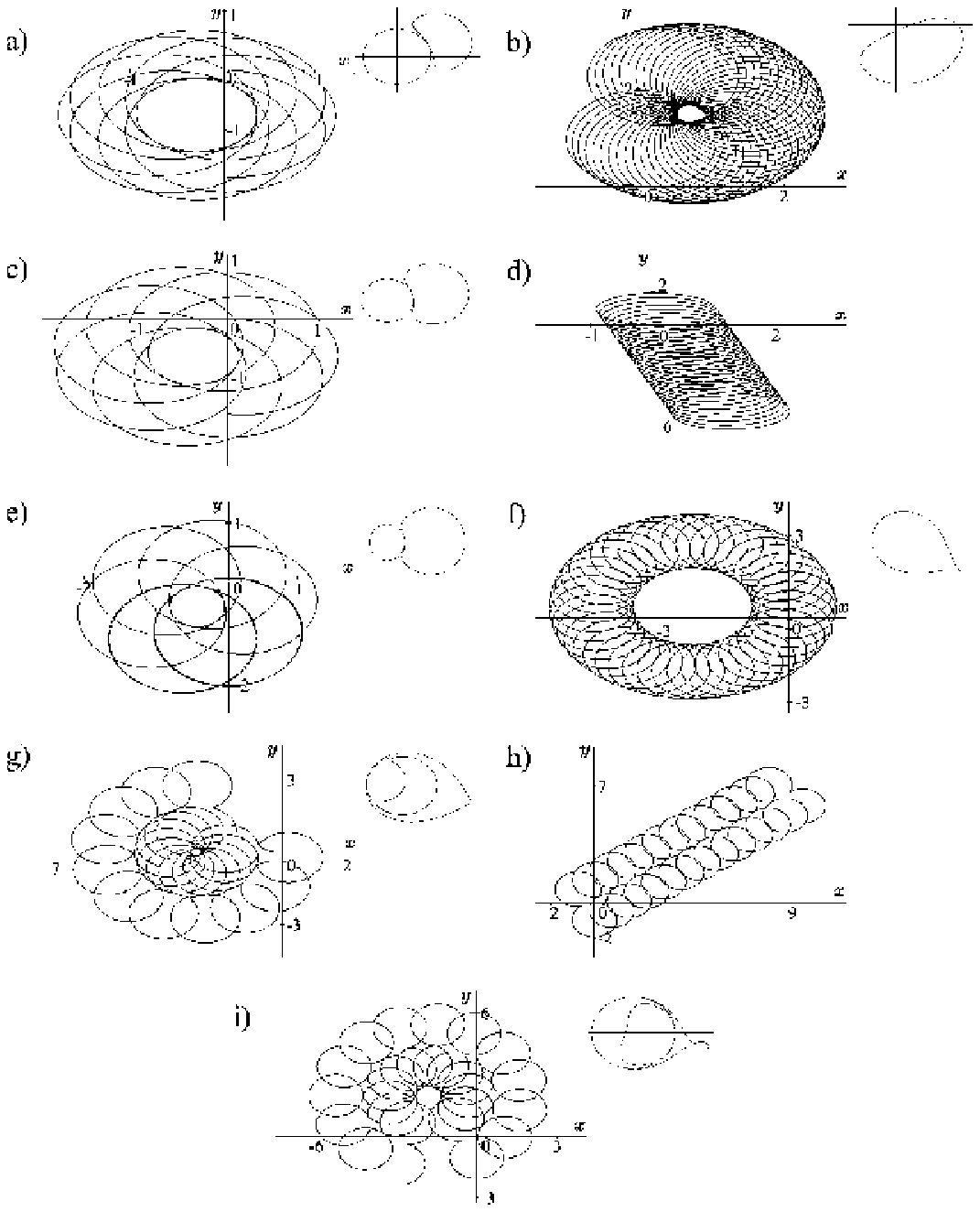} \caption{Trajectories of the point
of contact of the disk in the absolute space at various values of the
integral of energy. Parameters of the system correspond to figure a). The
closed trajectories in the frame of references rotating with the angular
velocity $\omega_{\psi}$ (see explanations in the text) are presented in
the right upper corner of each figure (except the case of infinite
motions). In figures a) and b) we present various types of motion of the
disk at the energy $h=0.86 $. Figures c) and d) correspond to the energy
$h=0.92217$ when one of the motions becomes resonant
($\omega_{\theta}=\omega_{\psi}^{(2)}$) and the secular drift (fig. d) is
observed. The increase of the energy in figures e) and f) to $h=0.961$
makes both types of motion bounded again. In figure g) the motion of the
disk is presented at $h=1.1$ after merging of two domains of possible
motions corresponding to various types of motion. The infinite motion in
figure h) corresponds to the resonance $\omega_{\psi}=2\omega_{\theta}$ at
the energy $h=1.18169$. In figure i) the motion of the point of contact of
the disk is presented after the further increase of the energy up to
$h=1.4$.} \label{pic6}
\end{center}
\end{figure}

\vspace{-2mm}

\subsection{The qualitative analysis of motion of apexes}
\vspace{-2mm}

The behavior of angles of proper rotation $\vfi$ and precessions $\psi$
that together with $\theta$ determined the motion of apexes is defined by
relations~(\ref{eq17}). The important feature in this problem is the {\it
two-frequency} behavior of each of these angles. That is not usual for
integrable systems. For example, for the Kovalevskaya top the angle
$\psi(t)$ is defined by three frequencies~\cite{BorMamDTT}. In this case
such phenomena is connected to the existence of two methods of reduction
with respect to the symmetries of system~(\ref{eq8}),~(\ref{eq912}).

\begin{figure}[!t]
\centerline{{\parbox{160mm}{%
\cfig<bb=0 0 159.4mm 90.7mm>{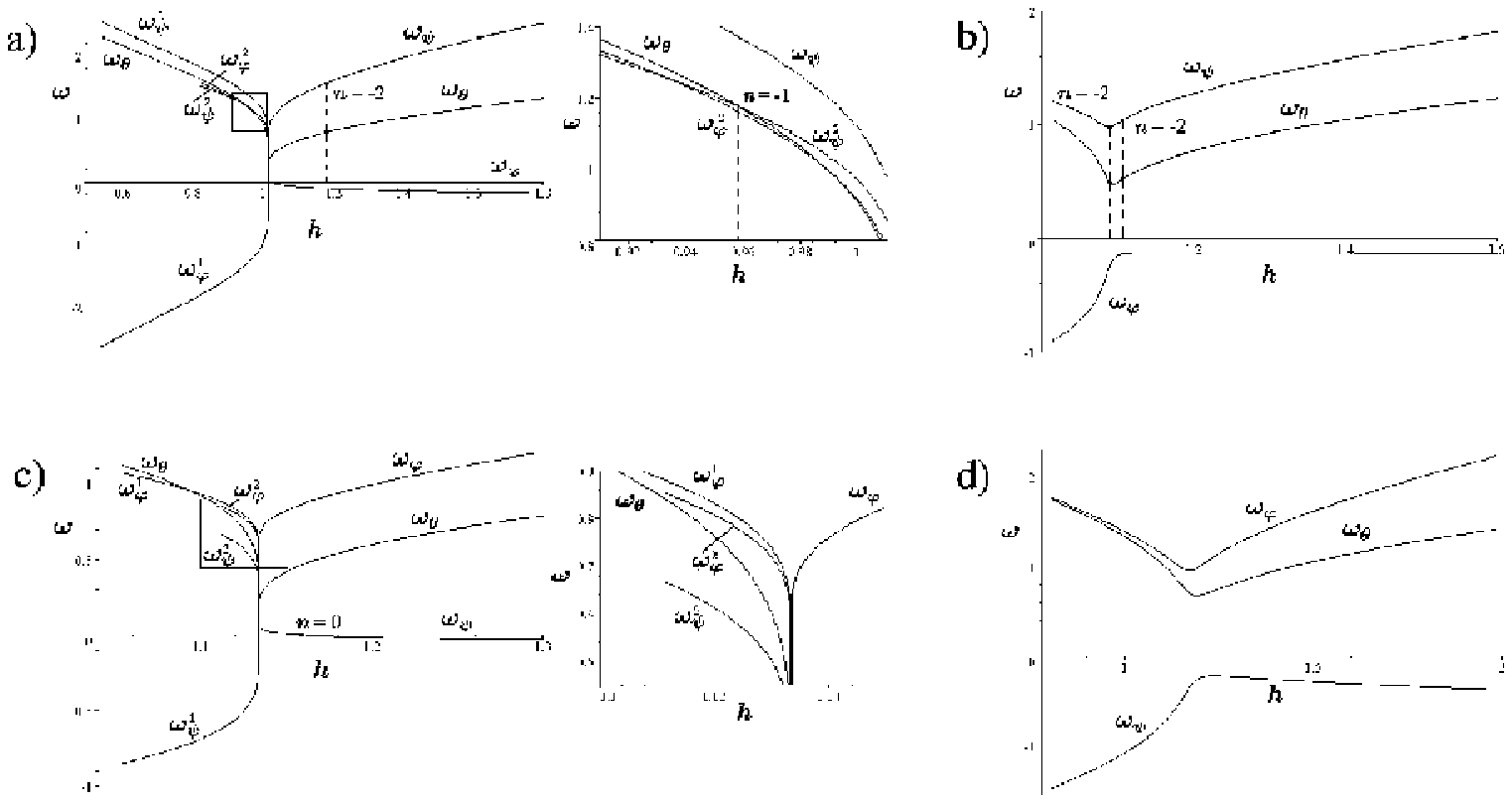} \caption{ Dependencies of
frequencies $\omega_{\theta},\,\omega_{\psi}$ and $\omega_{\vfi}$ from the
energy at $I_1 =\frac14,\,I_3=\frac12,\,R=1,\,m=1$, ${g=1}$, and various
values of integrals $C_1,\,C_2 $. In figures a) and c) the areas marked by
the rectangles are separately presented in the increased scale. In the
field of energies where two different types of motions are possible in the
absolute space we denote the frequencies by
$\omega_{\psi}^1,\,\omega_{\vfi}^1$ and
$\omega_{\psi}^2,\,\omega_{\vfi}^2$. The resonance energies are marked on
the graphs by thick dots. The orders of the resonances are indicated near
the dots. The values of integrals for the dependencies presented here are:
a) $C_1=0.04,\,C_2=-0.02$; \break b) $C_1=0.09,\,C_2=-0.07$;
c)$C_1=0.065,\,C_2=0.055$; d) $C_1=0.09,\,C_2=0.03$.}\label{pic7}}}}
\end{figure}

From the geometrical point of view whole space of variables
$\bM,\,\bs\alpha,\,\bs\beta,\,\bs\gamma$ is foliated on three-dimensional
tori defined as the joint level surfaces of the integrals $C_1,\,C_2,\,h$
and the geometrical integrals. The motion represents a winding of the
three-dimensional torus with frequencies
$\omega_{\theta},\,\omega_{\vfi},\,\omega_{\psi}$~\cite{Moshuk}. (For the
reduced systems~(\ref{eq8}) and~(\ref{eq912}) the corresponding tori are
two-dimensional.)

Since the frequencies depend only on constants of the first integrals, all
motion on the torus have the identical frequency that not evident for
nonholonomic systems. Even for the integrable nonholonomic systems on
two-dimensional tori there is a non-uniform rectilinear motion and,
generally speaking, the intermixing is possible (see paper~\cite{KozDAN}).

{\it Practically these arguments prove that the given system is
Hamiltonian one in the analytical sense (though the Hamiltonian function
can be a different from the energy \eqref{eq6} {\rm\cite{Moshuk}}.
Moreover, taking into account only the analytical point of view we can say
that near the nonsingular torus the system becomes the Hamiltonian one by
the infinite number of methods {\rm\cite{KozMulty}}.}

\begin{rem}
N.\,K.\,Moschuk in~\cite{Moschuk} observed a related phenomena studying
the nonholonomic Chaplygin system possessing some number of the linear
with respect to velocities first integrals.
\end{rem}

At the same time the existence of a natural (algebraic) Poisson structure
with a Hamiltonian defined by the energy~(\ref{eq6}) remains an open
problem. A.\,V.\,Borisov and I.\,S.\,Mamaev show that the reduced
system~(\ref{eq11}) is the Hamiltonian one with some algebraic
nonlinear bracket (see~\cite{BorMamInvMes}), 
however the possibility of its lifting on the systems~(\ref{eq8})
and~(\ref{eq912}) is still not investigated. \vspace{-2mm}

\subsection{The analysis of motion of the point of contact.}\vspace{-1mm}
{\advance\baselineskip by -.5pt
For the analysis of motion of the point of contact we decompose the
velocity~(\ref{eq18}) in the Fourier series with respect to time. Then
from~(\ref{eq17}) we get \nopagebreak\vspace{-2mm} \eqc*{
\dot{Z}=\smash{\sum\limits_{n\in\mathbb{Z}}}v_ne^{i(\omega_{\psi}+
n\omega_{\theta})t}. } Integrating with respect to time we
obtain\vspace{-2mm} \eqc*{
Z(t)=Z_0+e^{i\omega_{\psi}t}\sum\limits_{n\in\mathbb{Z}}\frac{v_n}{i(\omega_{\psi}+
n\omega_{\theta})}e^{in\omega_{\theta}t}. } Thus, if at $
\omega_{\psi}+n\omega_{\theta}\ne0$ we use the frame of references
rotating about the point $Z_0 $ with the angular velocity $\omega_{\psi}$,
then the point of contact traces some closed curve
(see~\cite{Moshuk},~\cite{KozKol}). Various types of such closed curves and
trajectories corresponding to them in the fixed space are presented in
figure~\ref{pic6}.

At the resonance $\omega_{\psi}+n\omega_{\theta}=0$ we observe the secular drift 
of the point of contact. Graphs of frequencies
$\omega_{\psi}(h),\,\omega_{\theta}(h),\,\omega_{\vfi}(h)$ at the fixed
values of integrals $C_1,\,C_2$ are presented in fig.~\ref{pic7}. They
show that the relation $\omega_{\psi}+n\omega_{\theta}=0$ can be fulfilled
both in the case of existence of one and of three regular precessions. And
at the same energy some initial conditions lead to a secular drift while
the others are not (see fig. \ref {pic6}). Since all frequencies depend
only on the values of the first integrals the relation
$\omega_{\psi}+n\omega_{\theta}=0 $ define in three-dimensional space of
integrals some two-dimensional manifold corresponding to the infinite
trajectories of the disk.

{\it Thus, for almost all initial conditions {\rm (}except the indicated
manifold{\rm)} all trajectories of the disk are bounded}.

We can consider this result to be opposite to the one obtained from
research of dynamics of the point of contact for the Chaplygin ball on a
horizontal plane (see~\cite{kilshar}) where the majority of trajectories,
on the contrary, were unbounded.
}

\end{document}